\documentclass{aa}
\usepackage{graphicx}
\def\12CO    {\hbox{\rm $^{12}$CO}}     
\def\13CO    {\hbox{\rm $^{13}$CO}}     
\def\ga {\lower.5ex\hbox{$\; \buildrel > \over \sim \;$}}
\def\la {\lower.5ex\hbox{$\; \buildrel < \over \sim \;$}}
\begin{document}
\title{HCN and HCO$^{+}$ emission in the disk of \object{M~31}}
\author{N. Brouillet\inst{1}
  \and S. Muller \inst{2,3}
  \and F. Herpin \inst{1}
  \and J. Braine \inst{1}
  \and T. Jacq \inst{1}}

 \institute{Observatoire de Bordeaux, Universit\'e Bordeaux 1- CNRS, B.P. 89, F-33270 Floirac, France.\\
 \email{name@obs.u-bordeaux1.fr}
\and
 IRAM, 300 rue de la Piscine, F-38406 Saint-Martin d'H\`eres, France. 
 \and
  Institute of Astronomy and Astrophysics,  Academia Sinica. P.O. Box 23-141, Taipei 106, Taiwan. \\
   \email{muller@asiaa.sinica.edu.tw}\\
  }

\offprints{Nathalie Brouillet, \email{brouillet@obs.u-bordeaux1.fr}}
\date{Received 19 September 2003 / Accepted 3 September 2004}
\abstract{We report observations made with the IRAM 30~m radiotelescope in the HCN(1--0) and HCO$^{+}$(1--0) lines towards a sample of molecular complexes (GMCs) in the disk of the Andromeda galaxy (\object{M~31}).
The targets were identified bright CO GMCs selected from the IRAM 30~m CO survey with various morphologies and environments. The clouds vary in galactocentric distances from 2.4 to 15.5~kpc. The HCN and HCO$^{+}$ emission is easily detected in almost all observed positions, with line widths generally similar  to the CO ones and there is a good correlation between the two dense gas tracers. The HCO$^{+}$ emission is slightly stronger than the HCN, in particular towards GMCs with a strong star formation activity. However the HCO$^{+}$ emission is weaker than the HCN towards a quiescent cloud in the inner part of \object{M~31}, which could be due to a lower abundance of  HCO$^{+}$. We derive ${\rm I_{\rm HCN}}/{\rm I_{\rm CO}}$ ratios between 0.008 and 0.03 and ${\rm I_{\rm HCO^{+}}}/{\rm I_{\rm CO}}$ ratios between less than 0.003 and 0.04. We study the radial distribution of the dense gas in the disk of \object{M~31}. Unlike our Galaxy the HCO$^{+}$/CO ratio is lower in the center of \object{M~31} than in the arms, which can be explained by both a lower abundance of  HCO$^{+}$ and different conditions of excitation. Furthermore the HCN/CO and HCO$^{+}$/CO ratios appear to be higher in the inner spiral arm and weaker in the outer arm.

 }

\maketitle
%

\section{Introduction}

The molecular gas in galactic centers has been investigated through
multi-molecule studies (e.g. Henkel et al.  \cite{henkel91}) but most
of our knowledge about the molecular gas in the disks of external galaxies
until now is only based on CO observations.  In our Galaxy a few key
molecules are used to obtain complementary information on the physical and
chemical state of the insterstellar medium and its radial distribution. 
HCN and HCO$^+$ are among these key molecules because their rotational line
emission is rather strong and, due to their large dipolar moment ($\mu $
= 2.98 and 4.48 Debye respectively) but similar frequencies, they are
excited in physical conditions different from those of CO ($\mu $ = 0.1
Debye).  In particular, the critical density to collisionally excite the 1--0 transitions of HCN
and HCO$^+$ is much higher ($n~\sim 10^{5-6}$~cm$^{-3}$) than that required
for the CO(1--0) transition ($n~\sim 10^3$~cm$^{-3}$).  Both molecules are linear and have
similar properties except the fact that HCO$^+$ being an ion is expected to
have a different chemical behaviour.  Because stars condense out of the densest material, the dense gas
tracers (e.g. HCN, HCO$^+$) may provide the best link with star formation. 
In IR-luminous galaxies, HCN is found to be much better correlated with the
star formation rate than the CO (Solomon et al. \cite{solomon}).  One can
wonder if a similar relationship exists within disks when the dense gas
tracers are observed.
 
With the much improved sensitivity of millimeter radiotelescopes, it is now
possible to observe the HCN and HCO$^+$ emissions towards external
galaxies.  Numerous studies have been made towards galactic nuclei (e.g.
Brouillet \& Schilke \cite{brouillet}; Chin et al.  \cite{chin}; Curran et
al.  \cite{curran}; Helfer \& Blitz \cite{helfera}; H\"{u}ttemeister et al. 
\cite{huettemeister}; Nguyen-Q-Rieu et al.  \cite{rieu92}; Paglione et al. 
\cite{paglione}; Seaquist \& Frayer \cite{seaquist}) but studies of
galactic disks are very few.  Braine et al.  (\cite{braine}) observed a few
positions in the HCN line along the major axis of \object{NGC 4414}.  HCN
emission is detected in the spiral arms, out to 3.5 kpc, but not in the interarms of
\object{M~51} (Kuno et al.  \cite{kuno}; Kohno et al.  \cite{kohno96}). 
Sorai et al.  (\cite{sorai}) recently observed HCN along the major axes of
six nearby CO bright galaxies, but out to 1kpc only.  Our goal is to extend our knowledge of the dense gas further out in galactic disks.

We have observed the HCN and HCO$^+$ emission in the disk of the closest large spiral
galaxy, \object{M~31}.  The recent high-resolution CO(1--0) survey made with the
IRAM 30~meter radiotelescope (Gu\'elin et al.  \cite{guelin}, Nieten et al. in prep.) allowed
identification of the Giant Molecular Clouds (GMCs) in \object{M~31} (spatial
resolution of 23$\arcsec$, i.e. $\sim$ 90~pc) and mapped the distribution
of these GMCs in the disk of the galaxy.  In addition, detailed CO(1--0) and CO(2--1)
observations have been performed with the IRAM Plateau de Bure Interferometer (PdBI) 
on a sample of GMCs with various morphologies in different environments and
located at galactic radii between 5 and 18~kpc (Neininger
\cite{neininger}, Muller \cite{muller}).  These GMCs, which have similar CO emission on the 90~pc
scale, reveal very different structures with the higher resolution of the
PdBI ($\sim$ 4~pc).  We have selected a few GMCs from the IRAM 30~m survey in order to detect HCN
and HCO$^+$ emission.

In this paper we present the results of our observations and we compare the
ratios between the 3 species HCN, HCO$^+$ and CO measured in the molecular
complexes with different physical properties and at different
galactocentric distances.  We study the fraction of dense molecular gas
(Sect.~\ref{dense}) and the radial distribution of the dense gas
(Sect.~\ref{radius}) in the disk of \object{M~31}, compared to our Galaxy and other
galaxies.

\section{Observations} \label{observations}

The adopted parameters for \object{M~31} are: center position 
$\alpha_{1950} = 00^{\rm h} 40^{\rm m} 00\,.\!\!^{\rm s}3$, 
$\delta_{1950} = 41\degr 00\arcmin 03\arcsec$ 
($\alpha_{2000} = 00^{\rm h} 42^{\rm m} 44\,.\!\!^{\rm s}32$,  
$\delta_{2000} = 41\degr 16\arcmin 08\,.\!\!{\arcsec}5$); an 
inclination with respect to the plane of the sky of 
$77\,.\!\!\degr5$; a distance of 780~kpc (Stanek \& Garnavich 
\cite{stanek}).

HCN(1--0) and HCO$^{+}$(1--0) line observations at 88.6~GHz and 89.2~GHz
were carried out towards \object{M~31} with the IRAM 30~m radiotelescope in
April, May and October 2002 and in June 2004.  The half power beamwidth was
$27\,.\!\!{\arcsec}5$, corresponding to a spatial resolution of 104~pc at
the distance of \object{M~31}.  The pointing was regularly checked on
nearby quasars and the absolute correction was $\leq 5\arcsec$.  In order to obtain flat baselines, we used a wobbler switching procedure with a reference position $4\arcmin$ offset from the source in azimuth. The CO(2--1) line was observed simultaneously to check that there was no emission at the same velocity in the reference position. We used two
SIS receivers in parallel when observing each line separately.  The system
temperatures ranged from 80 to 150~K. The temperatures were scaled in main
beam temperatures with values of the beam efficiency and the forward
efficiency at these frequencies of 0.77 and 0.95 respectively.  Two 1~MHz
filterbanks and an autocorrelator gave velocity resolutions of 3.4 and
1.1~km~s$^{-1}$ respectively.

\section{Results} \label{results}
 
\begin{figure*}
\centering
\includegraphics[width=16cm]{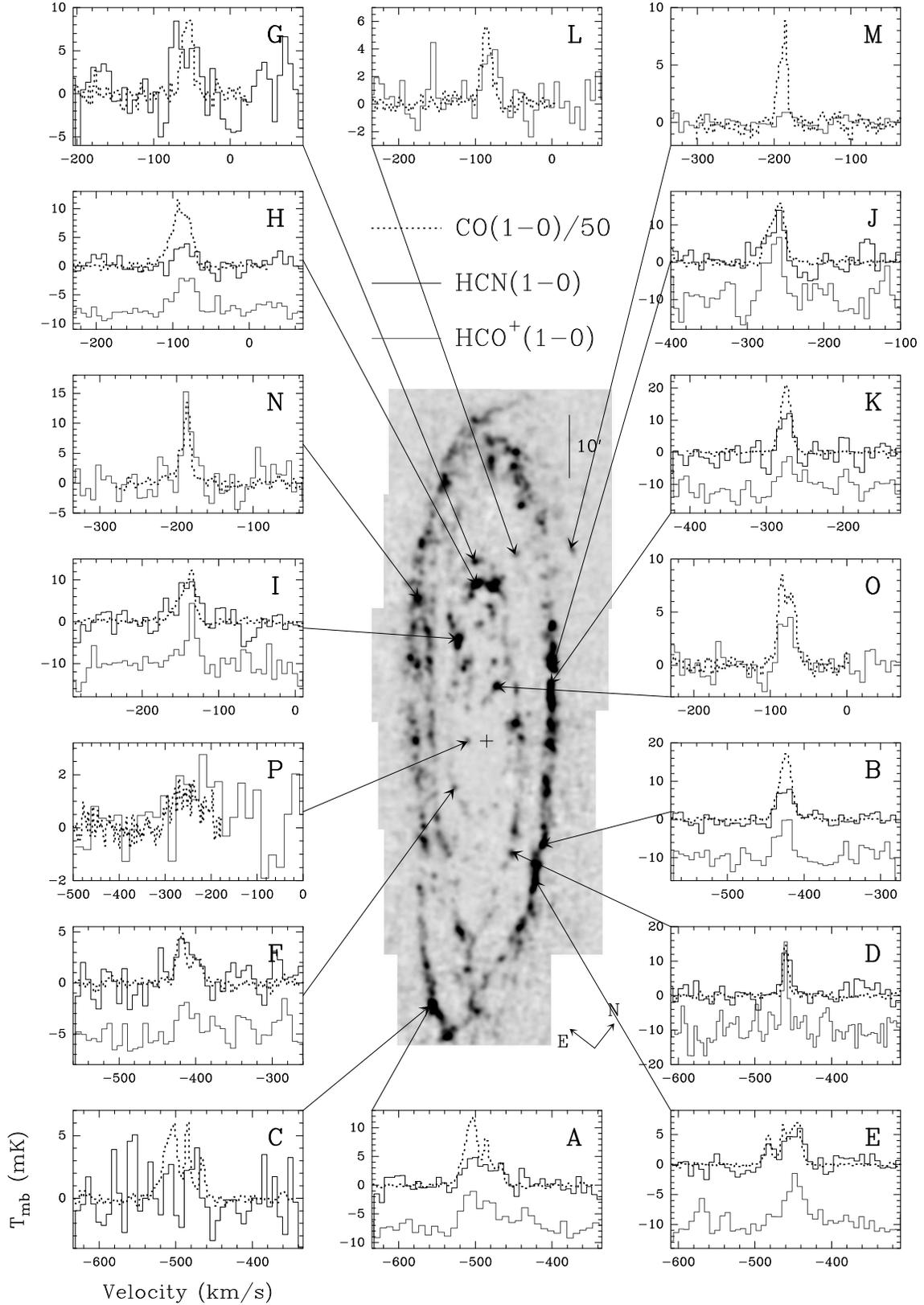}
\caption[] {The map is the CO(1--0) integrated intensity observed with 
the 30~m telescope and smoothed to a resolution of 
$45\arcsec$ (Nieten et al. in prep.). The arrows indicate the 
observed positions in the galaxy. The HCN(1--0), HCO$^{+}$(1--0) and CO(1--0) spectra are displayed; the HCO$^{+}$ spectra are shifted in 
intensities for a better comparison when the HCN spectra are present. The CO(1--0) spectra are convolved to 
the same spatial resolution as the HCN and HCO$^{+}$ observations, and the 
CO intensity is divided by 50.}
\label{spectra}
\end{figure*}

\begin{table}
\begin{center}
\caption[]{Column 1 is the name (letter) of the observed position followed, in column 2, by Hodge's number for the corresponding dark clouds (Hodge \cite{hodge}).  In column 3, 
X and Y are the offsets from the center along, respectively, the major and
minor axes of \object{M~31}.  Column 4 indicates the observed lines. 
Columns 5 to 7 are the spectral parameters derived from a gaussian fit of
the HCN and HCO$^{+}$ spectra: line velocity, half power line width and main beam
antenna temperature.  Column 8 is the rms of the spectra. The conversion factor from Kelvin (main beam scale) to Jansky for the 30~m radiotelescope at these frequencies is 4.7~Jy/K.}

\begin{tabular}{llr@{}r@{}llrrrc}
 \hline
 \multicolumn{2}{c}{Cloud} &
 \multicolumn{3}{c}{X,Y} &
 \multicolumn{1}{l}{Line} &
 \multicolumn{1}{c}{$V_{lsr}$} &
 \multicolumn{1}{c}{$\Delta$V} &
 \multicolumn{1}{c}{$T_{mb}$} &
 \multicolumn{1}{c}{$\sigma$} \\
 \multicolumn{1}{c}{} &
  \multicolumn{1}{c}{} &
 \multicolumn{3}{c}{$\arcmin$,$\arcmin$} &
 \multicolumn{1}{c}{} &
 \multicolumn{1}{c}{km/s} &
 \multicolumn{1}{c}{km/s} &
 \multicolumn{1}{c}{mK} &
 \multicolumn{1}{c}{mK}  \\
 \hline
A  & D39 & -41.9,&8.&6 & HCN & -499.5 & 31.5 & 4.7 & 2\\
 & & & & &  & -466.8 & 11.1 & 4.2 & 2\\
  & & & & &  HCO$^{+}$ & -505.5 & 19.1 & 7.1 & 1 \\ 
  & & & & &  &  -481.4 &  12.6 &  5.1 &  1 \\ 
B  & D84 & -16.6,&-8.&7 & HCN & -423.8 & 22.0 & 8.5 & 2\\
 & & & & & HCO$^{+}$ & -424.7 & 19.2 & 10.6 & 3\\
C  & D39 & -42.3,&8.&9 & HCN & & & & 3\\
D  & D153 & -17.7,&-4.&0 & HCN & -459.0 & 10.6 & 16 & 3\\
 & &  & & & HCO$^{+}$ & -460.0 & 6.0 & 31 & 7\\
E   &D47 & -22.2,&-7.&6 & HCN & -481.0 & 12.7 & 3.6 & 1 \\
 & & & & &  & -447.9 & 27.3 & 5.8 & 1 \\
 & & & & & HCO$^{+}$ & -447.6 & 28.9 & 7.5 & 2 \\
F  & D387 & -7.6,&5.&1 & HCN & -413.8 & 31.6 & 3.9 & 2 \\
  & & & & &  HCO$^{+}$ &  &  &  &  1.5  \\ 
G  & D630 & 28.8,&1.&9 & HCN & & & & 3 \\
H  & D615 & 25.0,&1.&6 & HCN & -83.8 & 22.6 & 5.0 & 2 \\
 & & & & &   HCO$^{+}$ &  -82.9 &   28.7 &   6.1 &  1 \\ 
I  & D573 & 16.3,&4.&5 & HCN & -140.0 & 32.2 & 9.1 & 2 \\
 & & & & & HCO$^{+}$ & -134.7 & 10.4 & 16.7 & 3 \\ 
J  & D400 & 11.5,&-10.&3 & HCN & -261.5 & 23.1 & 12 & 4 \\
 & & & & & HCO$^{+}$ & -264.4 & 24.3 & 16 & 5 \\ 
K  & D348 & 9.1,&-10.&0 & HCN & -273.9 & 16.6 & 13.5 & 4 \\
 & & & & & HCO$^{+}$ & -270.6 & 26.5 & 9.2 & 4 \\ 
L  &  D594 &  29.7,& -4.& 5 & HCO$^{+}$ &  -78.9 & 19.8 &  4.2 & 1 \\
M & D525 & 30.7,& -13.& 2 & HCO$^{+}$ &  &  &  & 0.5 \\
N  & D660 & 22.7,& 10.&6 &  HCO$^{+}$ &  -186.1 &  9.4 &  18.9 & 3 \\
O  & D478 & 8.7,& -1.& 5 & HCO$^{+}$ &-78.0 & 21.0 &  4.6 &  1 \\
P  & D451 & -0.1,&3.& 0 &  HCO$^{+}$ &  &  &  & 1 \\
 \hline
 \end{tabular}
\end{center}
\end{table}

We have observed 16 GMCs (noted from A to P) located in the spiral arms of 
\object{M~31}.  Figure~\ref{spectra} shows the position of the observed
GMCs in the galaxy as well as the HCN(1--0) and HCO$^{+}$(1--0) spectra displayed with the corresponding CO(1--0) spectra. 
The HCN and HCO$^{+}$ line parameters derived from the observations are
indicated in Table 1, and the HCN and HCO$^{+}$ integrated intensities in Table 2.  Four positions only (C, G, M and P ) are not detected; 11 positions were observed in HCN, 9 of them are detected;  14 positions were observed in
HCO$^{+}$, 11 of them are detected.  The non detected positions may only be due to a low signal-to-noise ratio in the spectra (positions G, M, P are at the limit of detection).

The sample of the observed regions consists of GMCs located in the inner spiral arm (R$\sim 5-7$~kpc): D, F, G, H, I, and in the outer spiral arm (R$\sim 9-13$~kpc): A, B, C, E, J, K, N; one GMC (L) is in the interarm and one GMC (M) is located in the outer disk at a large radius (15.5~kpc); two GMCs are near the center: P at 3.1~kpc and O (detected first by Allen \& Lequeux \cite{allen}) at 2.4~kpc. The sample covers different environments: isolated clouds (D, L), quiescent cloud (F), GMCs with signs of recent star formation like large HII regions (A, E, H, I, J, K, N). The GMCs display different structures as revealed by the single dish CO observations (narrow lines, multiple profiles) and by the interferometric CO observations for those observed with the PdBI (compact structures, several clumps, filaments). Our observations show that HCN and HCO$^{+}$ emission is easily detected in the GMCs in the inner and the outer spiral arms as well as in different kinds of molecular complexes; it is detected in the center and in the interarm where the CO emission is weak and it is marginally detected at a large radius.

Taking into account the noise, the line widths are very similar in HCN, HCO$^{+}$ and CO and the line profiles of the HCN and HCO$^{+}$ emissions as well.  This implies that
they trace gas from the same physical regions.  On the observed scale, the
line widths do not trace the gravitational potential of the galaxy but the
dynamics of the clouds.  The similarity of line widths for the different
molecules reflects the same velocity dispersion for the different emission
regions within the beam.  One possible explanation is a fractal scheme
where dense or moderately dense condensations are widespread in the complex. 
Helfer \& Blitz (\cite{helferb}) also find the same line widths and line
shapes for the HCN and CS features of their Milky Way plane survey, when
compared to the corresponding CO emission.  However, Pirogov
(\cite{pirogov}), in his sample of bright FIR sources, find that the HCN
and HCO$^{+}$ line widths are close to each other but lower than those of
CO.

For the  8 GMCs detected in both HCN and HCO$^{+}$, the HCO$^{+}$ integrated
emission is slightly stronger than the HCN emission, with an average of
I$_{\rm HCO^{+}}$~$\sim$ 1.2~I$_{\rm HCN}$.  Studies in our Galaxy and
external galaxies show that the overall HCN, HCO$^{+}$ and CO emission is
generally similar but with different line ratios and the I$_{\rm
HCO^{+}}$/I$_{\rm HCN}$ ratio seems to vary from galaxy to galaxy.  It is
slightly lower than 1 in our Galaxy where Pirogov (\cite{pirogov}) find a
mean value of 0.6$\pm$0.4, the ratio being a bit higher when data 
for dark clouds, small globules and cirrus cores are added to the bright FIR sources. 
He finds a strong correlation between the integrated intensities of HCN,
HCO$^{+}$, CS and NH$_{3}$, four molecules considered as high density tracers. 
Turner \& Thaddeus (\cite{turner}) also observe a good correlation between HCN
and HCO$^{+}$ brightnesses towards Galactic giant molecular clouds, the two
species having quite similar distributions;  they find that the line widths are often different but their observations are on a scale smaller than the GMCs. 
The HCN and HCO$^{+}$ integrated lines have also about the same strengths
in \object{NGC 4945} (ratio of 0.93, Henkel et al.  \cite{henkel90}) and in
\object{NGC 253} (ratio of 0.9, Nguyen-Q-Rieu et al.  \cite{rieu89}). 
However the I$_{\rm HCO^{+}}$/I$_{\rm HCN}$ ratio varies from 2 to less
than 0.25 in the sample of 15 galaxies observed by Nguyen-Q-Rieu et al. 
(\cite{rieu92}), the highest ratio corresponding to \object{M~82}.  A five
point map in \object{M~82} (Nguyen-Q-Rieu et al.  \cite{rieu89}) shows
that the ratio is 2 in the center and decreases down to 1.4
30$\arcsec$ further on.  In fact, when observing the center of
\object{M~82} on the molecular cloud scale (2$\arcsec$, i.e.
30~pc), the variations range from 0.75 to 3 depending on the molecular
clouds (Schilke \& Brouillet, unpublished).  Such a high
ratio (from 1.3 to more than 3.6) is also found by Chin et al. 
(\cite{chin}) towards HII regions in the Magellanic Clouds.  A high ratio
is consistent with a brighter HCO$^{+}$ emission due to the intense
ionization flux from supernovae and young stars, with a large extent of the
HCO$^{+}$ emission region and HCN coming from dense cloud cores.

HCN and HCO$^{+}$ are both linear molecules containing one atom of hydrogen and two "metals". Their dipole moments are quite similar, the frequencies of the transitions are almost identical, HCN and HCO$^{+}$ are excited collisionally by H$_{2}$ (Seaquist \& Frayer (\cite{seaquist})  show that no other mechanism is needed for the excitation of HCO$^{+}$ in the clouds of \object{M~82}), and if the abundances are estimated to be similar, it is not surprising that globally the HCN and HCO$^{+}$ line intensities are on the same order on the observed GMCs' scale. However variations in the I$_{\rm
HCO^{+}}$/I$_{\rm HCN}$ ratio exist from galaxies to galaxies and from GMCs to GMCs. These variations may be due to variations in density. For example, simple LVG calculations show that HCO$^{+}$ emission is thermalized at lower density than HCN, thus densities of 10$^5$~cm$^{-3}$ and column densities of roughly 10$^{15}$~cm$^{-2}$ may be sufficient to thermalize HCO$^{+}$, i.e. maximize the HCO$^{+}$ intensities, but are quite yet low for HCN. Observations of higher excitation lines are needed now to estimate the densities and the effect of their variations on the line ratios.

The variations in the HCO$^{+}$/HCN intensity ratio may also be due to variations in the abundance ratio due to differences in the ambiant radiation field. Unlike HCN, the abundance of HCO$^{+}$ varies with the cosmic-ray flux (the ionization of H$_{2}$ produces H$_{3}^+$ which reacts with CO to form HCO$^{+}$). This can explain the difference observed between galaxies (Nguyen-Q-Rieu et al.  \cite{rieu92}). \object{M~31} is a very quiet galaxy compared to \object{M~82} and we do not expect line ratios as high as in \object{M~82}. However the radiation field varies within the galaxy. Among the 9 GMCs observed in both HCN and HCO$^{+}$, 8 GMCs are detected in both lines. Seven are associated with large HII regions: positions A, B, E, H, I, J, K, most of them are near OB associations (Magnier et al.  \cite{magnier}) and we can suppose that the high star formation activity produces a strong ionization flux which can enhance the HCO$^{+}$ abundance. The  I$_{\rm HCO^{+}}$/I$_{\rm HCN}$ ratio is similar for these regions, from 1.1 to 1.5, except for position I where the ratio is apparently lower (but only one strong component is detected in HCO$^{+}$ and it is not clear if the difference for the second component is real or just due to the signal to noise ratio). The eighth GMC (D) is more isolated and presents a narrower line profile but an HII region is also located at the edge of the cloud and the HCO$^{+}$/HCN intensity ratio is similar (1.1). The only position not detected in both lines is F with I$_{\rm HCO^{+}}$/I$_{\rm HCN}\la 0.54$. F is located in the inner spiral arm, in the inner part of \object{M~31} where the star formation activity is weak; there is no HII region associated with F and the low ratio may be due to a difference of abundance. Anyway it is difficult to conclude as only one GMC of the sample of GMCs observed in both lines presents a clear different environment.

\begin{table*}
\begin{center}
\caption[]{The name of the observed positions and the galactocentric distances are indicated in columns 1 and 2 respectively.
Columns 3, 4, 5 display the integrated intensities of HCN(1--0), HCO$^{+}$(1--0) and
CO(1--0) emission, respectively.  Column 6 gives $\chi_{\rm N} =
100\times$I$_{\rm HCN}$/I$_{\rm CO}$, and column 7 $\chi_{\rm O^{+}} = 100\times$I$_{\rm
HCO^{+}}$/I$_{\rm CO}$.}

\begin{tabular}{lrrrrll}
\hline
 & R & I$_{\rm HCN(1-0)}$ & I$_{\rm HCO^{+}(1-0)}$ & I$_{\rm CO(1-0)}$ & 
 $\chi_{\rm N}$ & $\chi_{\rm O^{+}}$ \\
 & kpc & K km/s & K km/s & K km/s & & \\
\hline
A & 13.1 & 0.16 & 0.21 & 15.8 &  1.0 & 1.3 \\
  & & 0.05 & & 3.3 &  1.5 &  \\
B & 9.9 & 0.20 & 0.22 & 17.0 & 1.2 & 1.3 \\
C & 13.4 & $\la 0.11$* & & 10.2 &  $\la 1.1$ &  \\
D & 5.8 & 0.18 & 0.20 & 6.4 &  2.8 & 3.1  \\
E & 9.5 & 0.05 & & 2.6 & 1.9 &  \\
 & & 0.17 & 0.23 & 8.9 &  1.9 & 2.6 \\
F & 5.6 & 0.13 & $\la 0.07$* & 5.6 & 2.3 & $\la 1.3$  \\
G & 6.8 & $\la 0.15$* & & 7.7  & $\la 2.0$ & \\
H & 5.9 & 0.12 &  0.18 & 15.9 &  0.75 & 1.1 \\
I & 6.0 & 0.31 & 0.18 & 12.8 & 2.4 & 3.9**  \\
J & 11.1 & 0.30 & 0.42 & 19.5 & 1.5 & 2.15  \\
K & 10.7 & 0.24 & 0.26 & 17.7 & 1.35 & 1.5  \\
L & 8.2 &  & 0.09 & 4.3 &  & 2.1 \\
M & 15.5 &  &  $\la 0.015$* & 5.6 &  & $\la 0.3$  \\
N & 12.3 &  & 0.18 & 8.7 &  & 2.0 \\
O & 2.4 &  & 0.10 & 9.9 &  &1.0  \\
P & 3.1 &  &  $\la 0.16$* & 5.5 &  & $\la 2.9$ \\
\hline
\multicolumn{7}{l}{*Upper limits correspond to 3$\sigma \sqrt{{\Delta V_{line}}{\Delta V_{chan}}}$ 
where $\Delta V_{line}$ is the} \\
\multicolumn{7}{l}{CO(1--0) velocity width and $\Delta V_{chan}$ is the velocity resolution.} \\
\multicolumn{7}{l}{**Ratio of the components corresponding to the 
same velocity interval.}
\end{tabular}
\end{center}
\end{table*}

\section{Dense gas} \label{dense}

From the HCN, HCO$^{+}$ and CO(1--0) observations in \object{M~31} and other studies in our Galaxy and external galaxies, it appears that the HCN and HCO$^{+}$ integrated emissions are weakly correlated with the CO integrated emission on the same scale.  This implies a difference of dense gas content, or a difference of excitation or different HCN and HCO$^{+}$ abundances.

The ratio of HCN to CO integrated intensities is often used as a density
probe.  Helfer \& Blitz (\cite{helferb}) concludes that the I$_{\rm
HCN}$/I$_{\rm CO}$ ratio could indeed be a qualitative and even a
reasonable quantitative measure of pressure.  We define the 2 following ratios
$\chi_{\rm N} = 100\times$I$_{\rm HCN}$/I$_{\rm CO}$ and $\chi_{\rm O^{+}} =
100\times$I$_{\rm HCO^{+}}$/I$_{\rm CO}$. The values of $\chi_{\rm N}$ and $\chi_{\rm O^{+}}$ we derive for the GMCs in \object{M~31} are indicated in Table 2.

The ratio $\chi_{\rm N}$ varies from galaxy
to galaxy.  The numerous studies towards galactic centers lead to 
quite consistent values : 8 in our Galaxy (Jackson et al. 
\cite{jackson}), 8.4 in \object{NGC 4945} (Henkel et al. 
\cite{henkel90}), 11 in \object{NGC 6946} (Helfer \& Blitz \cite{helfera}),
2.9 for \object{M~82} and 6.9 for \object{NGC 253} (Nguyen-Q-Rieu et al. 
\cite{rieu89}), and from 8.6 to 60 in Seyfert galaxies (Kohno et al. 
\cite{kohno99}).  But due to the beam size and the distance to the galaxy,
the linear scale is often different and the value of the ratio is
completely different depending on the spatial resolution (see e.g. Jackson
et al.  \cite{jackson}).  It is thus difficult to compare the different
values and it is essential to identify the clouds to make real comparisons.

In the disk of our Galaxy, an average value of $\chi_{\rm N}=2.6\pm0.8$ is
found over size scales of 200~pc and $\chi_{\rm N}=1.4\pm2.0$ over nearby
individual GMCs on size scales of about 40~pc (Helfer \& Blitz
\cite{helferb}).  For the high-mass star forming regions observed by
Pirogov (\cite{pirogov}), the correlation between CO and HCN integrated
intensities is lower than between the high density tracers intensities,
with a mean value of $\chi_{\rm N}=9.8\pm8.2$.

The inner regions of a few external galaxies have been observed.  Downes et
al.  (\cite{downes}) mapped the center of \object{IC 342} and measured a
ratio $\chi_{\rm N}$ of $\sim$14 in the center and $\le$7 in the beginning
of the spiral arm.  Kuno et al.  (\cite{kuno}) detected a few positions in
the arms of \object{M~51} out to a radius of 3.5~kpc.  They found a ratio
of 1.1-1.9, much smaller than the value in the center of \object{M~51}
($\chi_{\rm N}$=8.6).
 
\begin{figure}
\includegraphics{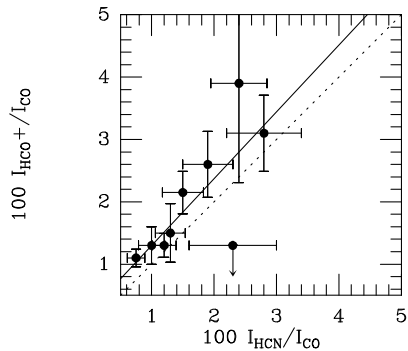}
\caption[]{$\chi_{\rm O^{+}}$ as a function of $\chi_{\rm N}$.  The dotted
line represents ${\rm I_{\rm HCO^{+}}}/{\rm I_{\rm CO}} = {\rm
I_{\rm HCN}}/{\rm I_{\rm CO}}$ and the solid line a least-square linear fit to the data ($ \chi_{\rm O^{+}} =  1.07 \chi_{\rm N} +0.23$) without taking into account position F (upper limit).}
\label{plot}
\end{figure}

Our observations present for the first time HCN and HCO$^{+}$ detections so
far out in a galactic disk, except our Galaxy.  The galactocentric distances of the observed GMCs range from  2.4 to 15.5~kpc (see Table 2).  We find mean values $\chi_{\rm N}=1.7\pm0.5$ and $\chi_{\rm
O^{+}}= 2.0\pm0.7$.  These values, derived on a size scale of 100~pc, are comparable to the one derived by  Helfer \& Blitz (\cite{helferb}) in the Galactic disk. Fig.~\ref{plot} represents $\chi_{\rm O^{+}}$ as a function of $\chi_{\rm N}$.  The upper limit under the ${\rm I_{\rm HCO^{+}}}/{\rm I_{\rm CO}} = {\rm
I_{\rm HCN}}/{\rm I_{\rm CO}}$ line corresponds to the quiescent GMC F (cf. end of Sect.~\ref{results}). If we consider only the GMCs with a star formation activity, there is a good correlation between the two
dense gas tracers with a slope of regression line of $\sim 1.1$ and ${\rm
I_{\rm HCO^{+}}}/{\rm I_{\rm CO}}$ is slightly higher than ${\rm I_{\rm
HCN}}/{\rm I_{\rm CO}}$. We can note that position I follows here the same trend as the other GMCs, as the  HCN and HCO$^{+}$ line intensities are divided by the CO line intensity of the component corresponding in velocity.

\section{Radial distribution of the dense gas} \label{radius}

A strong trend in the ratio of HCN to CO emission is present between the
nucleus and the disk of a galaxy but is uncertain within the disk.  For
example, the decrease of the ${\rm I_{\rm HCN}}/{\rm I_{\rm CO}}$ and ${\rm
I_{\rm HCO^{+}}}/{\rm I_{\rm CO}}$ ratios between the center and the disk
is observed in the inner regions of \object{M~51} and \object{NGC 253}
(Sorai et al.  \cite{sorai}; Rieu et al.  \cite{rieu89}).  Likewise, in our
Galaxy, ${\rm I_{\rm HCN}}/{\rm I_{\rm CO}}$ is about 0.08 averaged over the
central 600~pc and it is 0.026 in average over size scales of 200~pc between
R=3.5 and 7~kpc, and 0.014 over size scales of 40~pc in the solar
neighboorhood GMCs (Helfer \& Blitz \cite{helferb}).  A modest trend with
the radius within the plane of our Galaxy may also be found in the ratio of
HCO$^{+}$ to CO emission.  The HCO$^{+}$ emission peaks at R=4-5~kpc which
corresponds to the Galactic molecular ring, but the ratio HCO$^{+}$/CO
is almost constant (between $\sim$ 0.01 and $\sim$ 0.025), within the error
bars, from 3 to 9~kpc with a slight decrease at high radius (Liszt
\cite{liszt}).

\begin{figure}
\includegraphics{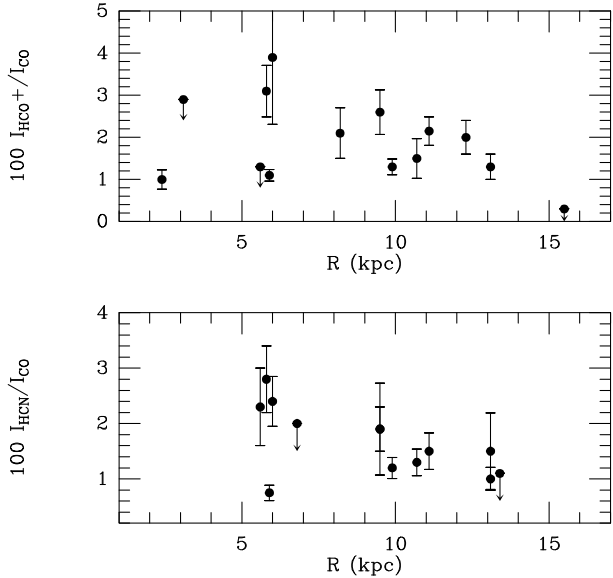}
\caption[]{Radial distributions of $\chi_{\rm N}$ and $\chi_{\rm O^{+}}$ in
the disk of \object{M~31}.}
\label{radial}
\end{figure}

Our data allows to investigate the distribution of the dense gas in the
disk of a galaxy, on a scale even larger than for our Galaxy.  We have
plotted the $\chi_{\rm N}$ and $\chi_{\rm O^{+}}$ ratios as a function of
the galactocentric radius R (Fig.~\ref{radial}). On the  $\chi_{\rm N}(\rm R)$ plot, we find a slight trend
similar as the one found in our Galaxy, the ratios decreasing with the
radius.  Position H (R = 5.9~kpc, $\chi_{\rm N}$= 0.75) is outside this tendency, but there seems to be a CO component in the spectrum which is not detected in HCN. As the ratios are
calculated for components at the same velocity, I$_{\rm HCN}$ could be
underestimated with respect to I$_{\rm CO}$ and the ratio could be a bit higher.  On the  $\chi_{\rm O^{+}}(\rm R)$ plot, we can find a similar trend as for $\chi_{\rm N}$ except for positions O (R = 2.4~kpc), F (R = 5.6~kpc) and H(R = 5.9~kpc). The lower value of $\chi_{\rm O^{+}}$ for F may be due to a lower HCO$^{+}$ abundance, and it could also be the case for position O but we have not the HCN observations to compare with. Loinard \& Allen (\cite{loinard}) confirmed the results of Allen \& Lequeux (\cite{allen}) that D478 (position O) is a large, massive, extremely cold molecular cloud. It is likely the case for the other dark clouds in the inner disk of  \object{M~31} where very little massive star formation is occurring, whereas the bright complexes in the arms of \object{M~31} are comparable in sizes and temperatures to Galactic GMCs. It is then possible that the HCO$^{+}$/HCN abundance ratio is lower in the center than in the arms where the GMCs are submitted to intense fluxes of UV photons. However even the $\chi_{\rm N}$ ratio may not increase towards the center, as measured in our Galaxy and other galaxies; we expect a higher ratio in the nucleus due to the higher densities and kinetic temperatures, but no molecular concentrations comparable to the Galactic ones are present near the center of  \object{M~31} (Gu\'elin et al.  \cite{guelin}) and the conditions of excitation appear to be very different. HCN observations would be needed to investigate the excitation of the dense gas in the center.

Concerning the values of $\chi_{\rm N}$ and $\chi_{\rm O^{+}}$ for the GMCs in the arms, we do not expect that the decrease with galactocentric distance is due to the general abundance gradient observed in spiral galaxies; the line integrated intensities do not drop with the radius but the line integrated intensities normalized by the CO intensities seem to decrease with R, and CO, HCN and HCO$^{+}$ all contain the same number of "rare" atoms.  The observed trend, if real (taking into account the uncertainties and position H which does not follow the tendency), would then be due to variations of excitation.
 
We can note that the GMC in the interarm (L at R = 8.2~kpc) presents a value of $\chi_{\rm O^{+}}$ comparable to the values in the arms. Although the GMC is isolated, the abundance of HCO$^{+}$ does not seem to be weak. We can consider that the chemistry was influenced by the conditions present while the cloud was in the arm; the abundances of the different species evolve during a period of about 10$^6$~years, then the chemistry remains stable as long as there is no perturbation, which is the case if the GMC goes out of the spiral arm and does not form any stars.
 
\section{Conclusion}

The HCN and HCO$^{+}$ emission is easily detected in our sample of GMCs spread over the disk of  \object{M~31} with a wide radial range (2.4-15.5~kpc) and with different structures and environments. The line widths and the line profiles are generally very similar in HCN, HCO$^{+}$ and CO. A velocity component is sometimes not detected in HCN or HCO$^{+}$, which may be due to the poor signal-to-noise ratio of the data.

The I$_{\rm HCO^{+}}$/I$_{\rm HCN}$ ratio of $\sim$ 1.2  observed in the disk of \object{M~31} is a bit higher than the one derived for most galaxies, including our Galaxy (ratio less than or equal to 1), but it is well below the values found in \object{M~82} and towards the HII regions in the Magellanic Clouds where the radiation fields are stronger.  However our sample of positions observed in both lines, consists mainly of GMCs associated with large HII regions and OB associations. The only position, situated in the inner part of  \object{M~31} and without any sign of star formation activity, is detected in HCN and marginally detected in HCO$^{+}$ with a line ratio $\la 0.54$. The I$_{\rm HCO^{+}}$/I$_{\rm HCN}$ ratio seems to be constant, within the uncertainties, for the GMCs in the arms with star formation activity, and the low ratio measured towards the quiescent cloud could be due to a lower abundance of HCO$^{+}$. 

The  $\chi_{\rm N} = 100\times$I$_{\rm HCN}$/I$_{\rm CO}$ ratio ranges from 0.75 to 2.8 and the $\chi_{\rm O^{+}} = 100\times$I$_{\rm HCO^{+}}$/I$_{\rm CO}$  ratio from less than 0.3 and 3.9. They are comparable to the values found in the Galactic disk and smaller than the values usually measured towards the center of galaxies.  We find low values of  $\chi_{\rm O^{+}}$ in the center of \object{M~31} compared to our Galaxy or galaxies like \object{M~51} and \object{NGC 253}, which can be due to a lower abundance of HCO$^{+}$ as the ionization field is weak in the inner part of \object{M~31}. But the conditions of excitation are also supposed to be different as there are no massive and warm GMCs in the nucleus as in the other galaxies.
When investigating the variation with the galactic radius in the \object{M~31} disk , we find that the $\chi_{\rm N}$ and $\chi_{\rm O^{+}}$ ratios are higher in the inner arm than in the outer arm. This weak trend, if real, is not supposed to come from the abundance gradient but from excitation effects.

\begin{acknowledgements}

We wish to thank Nikolaus Neininger for discussions and help while starting the project. We also thank the IRAM staff in Granada for their help during the observations.

\end{acknowledgements}


\begin{thebibliography}{}

\bibitem[1993]{allen} Allen, R.J., \& Lequeux, J. 1993, ApJ, 410, L15
\bibitem[1997]{braine} Braine, J., Brouillet, N., Baudry, A. 1997, A\&A, 
318, 19
\bibitem[1993]{brouillet} Brouillet, N., \& Schilke, P. 1993, A\&A, 277, 381
\bibitem[1997]{chin} Chin, Y.-N., Henkel, C., Whiteoak, J.B. et al. 1997, A\&A, 
317, 548 
\bibitem[2000]{curran} Curran, S.J., Aalto, S., Booth, R.S. 2000, A\&AS, 
141, 193
\bibitem[1992]{downes} Downes, D., Radford, S.J.E., Guilloteau, S., 
et al. 1992, A\&A, 262, 424
\bibitem[2000]{guelin} Gu\'elin, M. , Nieten, C., Neininger, N. et al. 2000, The interstellar medium in M31 and M33, proceedings 232, ed.  E. M. Berkhuijsen, R. Beck, and R. A. M. Walterbos.  Shaker, Aachen, 15
\bibitem[1997a]{helfera} Helfer, T.T., \& Blitz, L. 1997a, ApJ, 478, 162
\bibitem[1997b]{helferb} Helfer, T.T., \& Blitz, L. 1997b, ApJ, 478, 233
\bibitem[1990]{henkel90} Henkel, C., Whiteoak, J.B., Nyman, L.-\AA., Harju, 
J. 1990, A\&A, 230, L5
\bibitem[1991]{henkel91} Henkel, C., Baan, W.A., Mauersberger, R. 1991, 
A\&AR, 3, 47
\bibitem[1980]{hodge} Hodge, P.W. 1980, Atlas of the Andromeda Galaxy (Univ. of Washington Press)
\bibitem[1995]{huettemeister} H\"{u}ttemeister, S., Henkel, C., 
Mauersberger, et al. 1995, A\&A, 295, 571
\bibitem[1996]{jackson} Jackson, J.M., Heyer, M.H., Paglione, T.A.D., 
Bolatto, A.D. 1996, ApJ, 456, L91
\bibitem[1996]{kohno96} Kohno, K., Kawabe, R., Tosaki, T., Okumura, S.K. 
1996, ApJ, 461, L29
\bibitem[1999]{kohno99} Kohno, K., Kawabe, R., Vila-Vilar\'o, B. 1999, The 
Physics and Chemistry of the Interstellar Medium,in 3rd Cologne-Zermatt
Symposium, ed.  V. Ossenkopf, J. Stutzki, G. Winnewisser, 34
\bibitem[1995]{kuno} Kuno, N., Nakai, N., Handa, T., Sofue, Y. 1995, 
PASJ, 47, 745
\bibitem[1995]{liszt} Liszt, H.S. 1995, ApJ, 442, 163
\bibitem[1998]{loinard} Loinard, L., \& Allen, R.J. 1998, ApJ, 499, 227 
\bibitem[1993]{magnier} Magnier, A.E., Battinelli, P., Lewin, W.H.G. et al. 1993, A\&A, 278, 36
\bibitem[2003]{muller} Muller, S. 2003, Thesis
\bibitem[2000]{neininger} Neininger, N.  2000, The interstellar medium in M31 and M33, proceedings 232, ed.  E. M. Berkhuijsen, R. Beck, and R. A. M. Walterbos.  Shaker, Aachen, 25
\bibitem[1989]{rieu89} Nguyen-Q-Rieu, Nakai, N., Jackson, J.M. 1989, A\&A, 
220, 57
\bibitem[1992]{rieu92} Nguyen-Q-Rieu, Jackson, J.M., Henkel, C., Truong-Bach,
Mauersberger, R. 1992, ApJ, 399, 521
\bibitem[1997]{paglione} Paglione, T.A.D., Jackson, J.M., Ishizuki, S. 1997, ApJ, 484, 656 
\bibitem[1999]{pirogov} Pirogov, L. 1999, A\&A, 348, 600
\bibitem[2000]{seaquist} Seaquist, E.R., \& Frayer, D.T. 2000, ApJ, 540, 765
\bibitem[1992]{solomon} Solomon, P., Downes, D., Radford, S.J.E. 1992, ApJ, 387, L55
\bibitem[2002]{sorai} Sorai, K., Nakai, N., Kuno, N., Nishiyama, K. 2002, 
PASJ, 54, 179
\bibitem[1998]{stanek} Stanek, K.Z., \& Garnavich, P.M. 1998, ApJ, 503, L131
\bibitem[1977]{turner} Turner, B.E., \& Thaddeus, P. 1977, ApJ, 211, 755

\end{thebibliography}
\end{document}